\documentclass[aps,prd,twocolumn,superscriptaddress,showpacs,showkeys]{revtex4-1}
\usepackage{amsmath}
\usepackage{amsfonts}
 
\numberwithin{equation}{section}

\usepackage{graphicx,epstopdf}
\usepackage{pstricks,pst-plot,pst-grad,pst-3dplot,pstricks-add,pst-node}
\usepackage{mathrsfs}   
\usepackage{amssymb}    
\usepackage{CJKutf8}
\usepackage{CJK}
\usepackage{caption2,graphics,placeins,verbatim}
\usepackage{booktabs}

\begin{document}

\title{A scheme for neutrino mass generation through modified Higgs mechanism}
\author{Yanbin Deng$^{1}$}
\email{scientifichina@outlook.com, B305@cnu.edu.cn}
\author{Changyu Huang $^{2,3}$}
\email{cyhuang@purdue.edu}
\author{Yong-Chang Huang $^{4,5}$}
\email{ychuang@bjut.edu.cn}
\affiliation{$^{1}$ Department of Physics, Capital Normal University, Beijing 100048, China\\
$^{2}$Lawrence Berkeley National Laboratory, Berkeley CA 94720, USA\\
$^{3}$Department of Physics and Astronomy, Purdue University, W.Lafayette, IN 47907-2036, USA\\
$^{4}$Institute of Theoretical Physics, Beijing University of Technology, Beijing, 100022, China\\
$^{5}$Institute of Theoretical Physics, Jiangxi Normal University, Nanchang
Jiangxi 330022, China}
\date{\today }

\begin{abstract}
We propose a new method for neutrino mass generation through a modified Higgs mechanism by introducing an additional tiny vacuum breaking to the charged Higgs field. With identical particle spectrum as standard electroweak model, this modified Higgs mechanism endows the lepton-Higgs Yukawa coupling Lagrangian with a modified lepton mass matrix. Inserting the latest data of particle experiments into this modified lepton mass matrix enable us to locate the value of the newly introduced parameter parametrizing the extra perturbative Higgs vacuum breaking and produce masses for neutrinos of three generations. We reach a sum of three-generation neutrino massess at: $\Sigma_{(\nu)} ~ m_{\nu} \approx 0.104078 ~eV$. This prediction is supported by the result of the latest cosmological observation, i.e. the 95\% C.L. upper bound $\Sigma_{(\nu)} ~ m_{\nu} < 0.176 ~eV$ determined by the Planck data. In our proposal of the modified Higgs mechanism, the Higgs vacua break twice, such that, the very minuteness of the extra Higgs vacuum breaking explains the origin of the very minuteness of neutrino masses, while the relative greatness of the standard Higgs vacuum breaking is recognized as the origin of the relative greatness of charged lepton masses. This proposal can bring rich new physics to Higgs-relevant problems in particle physics, besides the massive neutrino problems.
\end{abstract}
\maketitle


\section{Introduction}

Neutrinoes have been one type of most concerned objects among fundamental particles for various reasons such as their critical role in the evolution of universe and their thus far unexplained puzzle of the gerneration of their masses. \cite{D. Kazanas, Z.Z.Xing, Lawrence M. Krauss, Guillermo Ballesteros} In the standard electroweak model of particle physics, neutrinos are considered as massless particles. Over the decades, neutrino oscillations had been experimentally discovered and confirmed by several collaborations \cite{Fukuda, Ashine, Ahn, Michael, Davis, Hampel, Altman, Abdurashitov, Ahmad, Eguchi, Abe, An}. The implication of these experiments directs to the flavor mixing between different generations of neutrinos thus the existence of nonzero neutrino masses \cite{Wolfenstein, Mikheyev, Fritzsch, King1, Altarelli, Mohapatra1, Gonzalez-Garcia, King2}.

Various mechanisms have been constructed to explain the fermion mass generation and flavor mixing and can roughly be classified into four types: (a) radiative mechanisms \cite{Weinberg1, Zee, Chao-Qiang Geng, Kazuki Enomoto}, (b) texture zeros \cite{Fritzsch1, Weinberg2, Wilczek, Fritzsch2}, (c) family symmetries \cite{Harari, Froggatt}, and (d) seesaw mechanisms \cite{Fritzsch3, Minkowski, Yanagida, Gell-Mann, Mohapatra, S. M. Barr, J. Schechter}. However, different theoretical schemes are plagued with various drawbacks, such as introducing new particle components that possess extraordinarily obscure prospects for future experimental tests, say the super heavy neutrino required in the seesaw mechanism.

In this work we propose a new scheme for the physics of massive neutrinos through a modification of the standard Higgs mechanism in which the vacuum expectation of the Higgs doublet is set to break the $SU_L(2)\times U_Y(1)$ gauge symmetry to bring the theoretical Glashow-Weinberg-Salam model down to real physics. This theme of study continued to evolved into composite Higgs which is supposed to be a bound state of massive strongly interacting fermions. \cite{David B. Kaplan} The Higgs particle in the Gell–Mann-Levy linear sigma model may also be a composite object as it turned out in QCD. \cite{YAMAWAKI} A bunch of progresses have been made in this direction over the years. \cite{Kaustubh Agashe, Charalampos Anastasiou, Ben Gripaios, Brando Bellazzini} In the standard Higgs mechanism, only the neutral Higgs field is assigned with a nonzero VEV. In this work, based on the standard Higgs mechanism which breaks the electroweak gauge symmetry by letting the neutral Higgs field component to assume nonzero VEV, we continue to introduce an additional tiny valued VEV for the charged Higgs field component. It can be shown that due to the introduction of this tiny valued charged Higgs VEV, even if working with the identical particle constituents as standard electroweak model, extra terms not present in the standard Higgs mechanism will appear in the modified lepton-Higgs Yukawa coupling Lagrangian. This gives the modified lepton mass matrix from which we can produce relations between neutrino masses and the masses of charged leptons and the tiny perturbative cacuum breaking parameter. By using current neutrino experimental data, we can determine the value of the extra perturbative Higgs vacuum breaking parameter and predict the masses for neutrinos of three generations, satisfying all the constraints of current particle experiments and cosmological observations.

The article is organized as follows: in Sec. 2, we layout the modified Higgs mechanism, and with identical particle constituents as standard model, derive the modified Lagrangian for lepton-Higgs Yukawa coupling carrying the modified lepton mass matrix. In Sec. 3, we solve the modified lepton mass matrix for mass eigenvalues, eigenstates and construct the lepton flavor-mass transformation matrix. In Sec. 4, we derive the dependence of neutrino masses on the masses of charged leptons and the tiny valued charged Higgs VEV. In Sec. 5, we work out the value of the tiny charged Higgs VEV using the latest data of neutrino experiments, and employ this result to produce the value of the neutrino masses of three generations. Sec. 6 is the conclusion and discussion.

\section{Lepton-Higgs Yukawa coupling in modified Higgs mechanism}

In the standard electroweak model, the breaking of the $SU_{L}(2)\times U_{Y}(1)$ symmetry for the gauge fields is induced by the Higgs field assuming the standard vacuum expectation value, $\Phi _{VEV}=\frac{1}{\sqrt{2}}\left\vert
\begin{array}{c}
0 \\
\upsilon
\end{array}
\right\vert $, for the intact massless gauge fields to realize as physically massive W and Z bosons and massless photons \cite{Englert, Higgs1, Higgs2, Guralnik, Higgs3, Kibble, Peskin}. We now study the new physics introduced by the assumption of a very tiny perturbation $\upsilon _{+}$ to the Higgs field vacuum expectation,
\begin{equation}
\Phi _{VEV}=\frac{1}{\sqrt{2}}\left\vert
\begin{array}{c}
\upsilon _{+} \\
\upsilon
\end{array}
\right\vert ,\upsilon _{+}\rightarrow 0  \label{2.1}
\end{equation}
where $\upsilon _{+}\rightarrow 0$ (but $\neq 0$) may be taken as a perturbation at the traditional zero point (which ensures electric charge conservation \cite{pdg2018}) and extremely small so that it fits all the current theories having been proved by the current experiments, i.e., the magnitude of which should be constrained by all the current particle experiments and cosmological observations \cite{pdg2018}.

In the standard electroweak model, when the vacuum breaking takes effect, i.e., when $\Phi _{VEV}=\frac{1}{\sqrt{2}}\left\vert
\begin{array}{c}
0 \\
\upsilon
\end{array}
\right\vert $ is being inserted into, e.g., the Lagrangian of the field system $\mathcal{L}_{H}=(D_{\mu }\Phi )^{+}D^{\mu }\Phi -V(\Phi )$, the orignal $SU_L(2)\times U_{Y}(1)$ gauge symmetry of the system would break to only the $U_{em}(1)$ symmetry for the eletromagnetic sector of the system, i.e., the system after vacuum breaking is deprived of the $SU(2)$ symmetry \cite{pdg2018, Peskin}. After these procedures have been taken out, then for the system deprived of $SU(2)$ symmetry, we introduce an extra tiny breaking to the charged component of the Higgs doublet, treat it as purely mathematical perturbation, and await for the new physics it can bring to us. This is the idea of doubly-vacuum-breaking Higgs mechanism. In the following work we show that this mofied Higgs mechanism when applied to the sector of neutrinos can produce very desirable results for neutrino masses. Also it can be shown that our proposal of an apparently QED breaking assumption causes only negligible correction to the photon mass, safely staying well below the constraint determined by latest experiments.

Then, playing the role of the Higgs field in standard Higgs mechanism, is the modified Higgs field with mathematical perturbation,
\begin{equation}
\Phi =\frac{1}{\sqrt{2}}\left\vert
\begin{array}{c}
\upsilon _{+} \\
\upsilon +H
\end{array}
\right\vert.  \label{2.2}
\end{equation}
Putting this perturbed Higgs field, back into exactly the identical procedure as in the standard Higgs mechanism, we obtain the Lagrangian for lepton-Higgs Yukawa coupling,
\begin{eqnarray}
&& \mathcal{L}_{\xi Yukawa} = -g_{\xi }\bar{l}_{\xi R}\Phi^{\dagger}l_{\xi
L}+h.c. \notag\\
& = & -g_{\xi }\bar{\xi }_{R}\frac{1}{\sqrt{2}}(
\begin{tabular}{ll}
$\upsilon _{+}^{\dagger}$ & $\upsilon +H$
\end{tabular}
)\left(
\begin{tabular}{l}
$\nu _{\xi L}$ \notag\\
$\xi _{L}$
\end{tabular}
\right) \notag\\
&& -g_{\xi }(
\begin{tabular}{ll}
$\bar{\nu }_{\xi L}$ & $\bar{\xi }_{L}$
\end{tabular}
)\frac{1}{\sqrt{2}}\left(
\begin{array}{c}
\upsilon_{+} \notag\\
\upsilon + H
\end{array}
\right) \xi _{R}\\
& = & -\frac{1}{\sqrt{2}}g_{\xi } \left[\bar{\xi }_{R}[\upsilon _{+}^{\dagger}\nu_{\xi L} + (\upsilon +H)\xi _{L}]\right. \notag\\
&& + \left. [\bar{\nu }_{\xi L}\upsilon_{+} + \bar{\xi }_{L}(\upsilon +H)]\xi _{R}\right] \notag\\
& = & -\frac{1}{\sqrt{2}}g_{\xi }\left[\upsilon_{+}^{\dagger}\bar{\xi }_{R}\nu
_{\xi L}+(\upsilon +H)\bar{\xi}_{R}\xi _{L}) \right. \notag\\
&& + (\bar{\nu}_{\xi L}\xi_{R}\upsilon_{+} + \bar{\xi}_{L}(\upsilon +H)\xi _{R}] \notag\\
& = & -\frac{1}{\sqrt{2}}g_{\xi }\left[\upsilon_{+}^{\dagger}\bar{\xi }_{R}\nu_{\xi L} + (\upsilon +H)\xi^{\dagger}\frac{1+\gamma _{5}}{2}\gamma _{0}\frac{1 - \gamma_{5}}{2}\xi \right. \notag\\
&& \left. + \bar{\nu }_{\xi L}\xi _{R}\upsilon_{+} + \bar{\xi }_{L}(\upsilon +H)\xi _{R}\right] \notag\\
& = &-\frac{1}{\sqrt{2}}g_{\xi }\left[\upsilon_{+}^{\dagger}\bar{\xi }_{R}\nu _{\xi
L}+(\upsilon +H)\xi^{\dagger}\gamma_{0}\frac{1-\gamma_{5}}{2}\xi \right.  \notag\\
&& \left. + \bar{\nu }_{\xi L}\xi _{R}\upsilon_{+}+\bar{\xi }_{L}(\upsilon +H)\xi
_{R}\right] \notag\\
& = & -\frac{g_{\xi}}{\sqrt{2}}\left[(\upsilon_{+})^{\dagger}\xi \nu _{\xi L}+\upsilon_{+}\bar{\nu}_{\xi L}\xi \right] - \frac{g_{\xi }\upsilon}{\sqrt{2}}\bar{\xi}\xi \label{2.3}\\
&& -\frac{g_{\xi }}{\sqrt{2}}H\bar{\xi }\xi, ~(\xi =e, \mu, \tau), \notag
\end{eqnarray}
where $g_{\xi }$ is the Yukawa coupling constant for the $\xi$-th generation leptons, and $l_{\xi R}$ and $l_{\xi L}$ are, respectively, the right-hand singlet and left-hand doublet of the $\xi$-th generation lepton.

Because we assume $\upsilon _{+}\rightarrow 0$ to be mathematical perturbation, we do not expect drastic modification on the properties of massive particles, then we respect the mass of the $\xi$-th generation charged lepton of the standard Higgs vacuum breaking,
\begin{equation}
m_{\xi} = \frac{g_{\xi}}{\sqrt{2}}\upsilon , ~ (\xi =e,\mu ,\tau).  \label{2.4}
\end{equation}
Substituting Eq.(\ref{2.4}) into Eq.(\ref{2.3}), we have
\begin{eqnarray}
\mathcal{L}_{Yukawa} & = & -\frac{m_{\xi}}{\upsilon}\left[(\upsilon_{+})^{\dagger}
\bar{\xi}\nu_{\xi L} + \upsilon_{+}\bar{\nu }_{\xi L}\xi \right] - m_{\xi}\bar{\xi }\xi \notag\\
&& - \frac{m_{\xi}}{\upsilon}H\bar{\xi}
\xi \notag\\
& = & -\left(
\begin{array}{cc}
\bar{\nu}_{\xi L} & \bar{\xi}
\end{array}
\right)
\left(
\begin{array}{cc}
0 & \frac{\upsilon _{+}}{\upsilon }m_{\xi } \\
\frac{(\upsilon _{+})^{\dagger}}{\upsilon }m_{\xi } & m_{\xi }
\end{array}
\right) \left(
\begin{array}{c}
\nu _{\xi L} \\
\xi
\end{array}
\right) \notag\\
&& -\frac{m_{\xi }}{\upsilon }H\bar{\xi }\xi, ~ (\xi =e,\mu ,\tau). \label{2.5}
\end{eqnarray}
It is the Yukawa coupling Lagrangian for three generations of leptons with mathematically perturbed Higgs field carrying modified mass matrix from standard model.

\section{Lepton mass eigenvalues, eigenstates and flavor-mass representation transformation}

The eigenvalues of the lepton mass matrix in Eq.(\ref{2.5}) (denoting it by $A$) can be computed as follows,
\begin{equation}
\det (A-\lambda )=\det \left(
\begin{array}{cc}
-\lambda & \frac{\upsilon _{+}}{\upsilon }m_{\xi } \\
\frac{(\upsilon_{+})^{\dagger}}{\upsilon }m_{\xi } & m_{\xi }-\lambda
\end{array}
\right) =0.  \label{3.1}
\end{equation}
Eq.(\ref{3.1}) can be expanded out as the quadratic equation,
\begin{equation}
\lambda ^{2}-\lambda m_{\xi }-\frac{(\upsilon _{+})^{\dagger}\upsilon _{+}}{
\upsilon ^{2}}m_{\xi }^{2}=0,  \label{3.2}
\end{equation}
with the following solutions,
\begin{eqnarray}
\lambda_{\pm } &=&\frac{m_{\xi }}{2}\left(1\pm \sqrt{1+4\frac{(\upsilon _{+})^{\dagger}\upsilon _{+}}{\upsilon ^{2}}}\right).  \label{3.3}
\end{eqnarray}
Thus the Lagrangian of Eq.(\ref{2.5}) can be written in diagonal form in the mass eigenstate representation,
\begin{eqnarray}
\mathcal{L}_{Yukawa} &=&-\left(
\begin{array}{cc}
\bar{\nu }_{\xi L}^{\prime } & \bar{\xi }^{\prime }
\end{array}
\right) \left(
\begin{array}{cc}
\lambda _{-} & 0 \\
0 & \lambda _{+}
\end{array}
\right) \left(
\begin{array}{c}
\nu _{\xi L}^{\prime } \\\
\xi ^{\prime }
\end{array}
\right) \notag \\
&& -\frac{m_{\xi }}{\upsilon }H\bar{\xi }\xi \notag\\
&=& -\frac{m_{\xi }}{2}\left(1-\sqrt{1+4\frac{(\upsilon _{+})^{\dagger}\upsilon _{+}}{
\upsilon ^{2}}}\right)\bar{\nu }_{\xi L}^{\prime }\nu _{\xi L}^{\prime } \notag\\
&& -\frac{m_{\xi }}{2}\left(1 +\sqrt{1+4\frac{(\upsilon _{+})^{\dagger}\upsilon _{+}}{\upsilon ^{2}}}\right){\bar{\xi}}^{\prime} \xi^{\prime} \label{3.4}\\
&& -\frac{m_{\xi }}{\upsilon }H\bar{\xi } \xi , ~ (\xi =e,\mu ,\tau). \notag
\end{eqnarray}
To find the mass-state eigenvectors, using Eq.(\ref{3.1}),
\begin{equation}
\left(
\begin{array}{cc}
-\lambda & \frac{\upsilon _{+}}{\upsilon }m_{\xi } \\
\frac{(\upsilon _{+})^{\dagger}}{\upsilon }m_{\xi } & m_{\xi }-\lambda
\end{array}
\right) \left(
\begin{array}{c}
a \\
b
\end{array}
\right) =0,  \label{3.5}
\end{equation}
expanding out as,
\begin{equation}
-\lambda a+\frac{\upsilon _{+}}{\upsilon }m_{\xi }b=0,~\left(\Rightarrow a=\frac{\upsilon_{+}}{\upsilon \lambda }m_{\xi }b \right),  \label{3.6}
\end{equation}
\begin{equation}
\frac{(\upsilon_{+})^{\dagger}}{\upsilon}m_{\xi}a+(m_{\xi}-\lambda )b=0.
\label{3.7}
\end{equation}
The eigenvector upto a constant is,
\begin{equation}
V=\left(
\begin{array}{c}
\frac{\upsilon _{+}}{\upsilon \lambda _{\pm }}m_{\xi }b \\
b
\end{array}
\right).  \label{3.8}
\end{equation}
By the normalization condition of eigenvector,
\begin{eqnarray}
V^{\dagger}V &=& \left(
\begin{array}{c}
\frac{\upsilon _{+}}{\upsilon \lambda _{\pm }}m_{\xi }b \\
b
\end{array}
\right)^{\dagger}
\left(
\begin{array}{c}
\frac{\upsilon _{+}}{\upsilon \lambda _{\pm }}m_{\xi }b \\
b
\end{array}
\right)  \notag \\
&=& \left(\frac{\upsilon_{+}}{\upsilon \lambda_{\pm}}m_{\xi }b \right)^{\dagger}\left(\frac{\upsilon_{+}}{\upsilon \lambda_{\pm}}m_{\xi}b \right) + b^{\dagger}b = 1.  \label{3.9}
\end{eqnarray}
From Eq.(\ref{3.9}), we find
\begin{equation}
b_{\lambda _{\pm }} = \frac{e^{i\theta _{\lambda _{\pm }}}}{\sqrt{1+\frac{\left\vert \upsilon_{+}\right\vert ^{2}}{\upsilon^{2}}\frac{m_{\xi }^{2}}{\lambda_{\pm }^{2}}}},  \label{3.10}
\end{equation}
where $\theta_{\lambda_{\pm}}$ are two arbitrary real phases. Finally we obtain the mass-state eigenvector,
\begin{equation}
V_{\lambda _{\pm }}=\left(
\begin{array}{c}
\frac{\upsilon _{+}}{\upsilon \lambda _{\pm }}m_{\xi }\frac{e^{i\theta
_{\lambda _{\pm }}}}{\sqrt{1+\frac{\left\vert \upsilon _{+}\right\vert ^{2}}{
\upsilon ^{2}}\frac{m_{\xi }^{2}}{\lambda _{\pm }^{2}}}} \\
\frac{e^{i\theta _{\lambda _{\pm }}}}{\sqrt{1+\frac{\left\vert \upsilon
_{+}\right\vert ^{2}}{\upsilon ^{2}}\frac{m_{\xi }^{2}}{\lambda _{\pm }^{2}}}
}
\end{array}\right).  \label{3.11}
\end{equation}
Using Eq.(\ref{3.11}), we achieve the transformation matrix between the flavor state and the mass state,
\begin{eqnarray}
U= && \left(
\begin{array}{cc}
\frac{\upsilon _{+}}{\upsilon \lambda _{-}}m_{\xi }\frac{e^{i\theta_{\lambda _{-}}}}{\sqrt{1+\frac{\left\vert \upsilon _{+}\right\vert ^{2}}{\upsilon ^{2}}\frac{m_{\xi }^{2}}{\lambda _{-}^{2}}}} & \frac{\upsilon _{+}}{\upsilon \lambda _{+}}m_{\xi }\frac{e^{i\theta _{\lambda _{+}}}}{\sqrt{1 + \frac{\left\vert \upsilon _{+}\right\vert ^{2}}{\upsilon ^{2}}\frac{m_{\xi}^{2}}{\lambda _{+}^{2}}}} \\
\frac{e^{i\theta _{\lambda _{-}}}}{\sqrt{1+\frac{\left\vert \upsilon_{+}\right\vert ^{2}}{\upsilon ^{2}}\frac{m_{\xi }^{2}}{\lambda _{-}^{2}}}} & \frac{e^{i\theta _{\lambda _{\pm }}}}{\sqrt{1+\frac{\left\vert \upsilon_{+}\right\vert ^{2}}{\upsilon ^{2}}\frac{m_{\xi }^{2}}{\lambda_{+}^{2}}}}
\end{array}
\right)\label{3.12}\\
= && \frac{1}{\sqrt{1+\frac{\left\vert \upsilon _{+}\right\vert ^{2}}{\upsilon
^{2}}}}\left(
\begin{array}{cc}
\frac{\upsilon _{+}}{\upsilon }e^{i\theta _{\lambda _{-}}} & \frac{\upsilon
_{+}}{\upsilon }e^{i\theta _{\lambda _{+}}} \\
\frac{e^{i\theta _{\lambda _{-}}}}{\sqrt{\frac{4}{ \left( 1 -\sqrt{1 + 4\frac{\left\vert \upsilon_{+}\right\vert ^{2}}{\upsilon ^{2}}} \right)^{2}}}} & \frac{e^{i \theta_{\lambda_{+}}}}{\sqrt{\frac{4}{\left( 1 + \sqrt{1 + 4 \frac{\left\vert \upsilon_{+}\right\vert ^{2}}{\upsilon ^{2}}} \right)^{2}}}}
\end{array} \right).  \notag
\end{eqnarray}
Transforming the flavor state to mass state,
\begin{eqnarray}
&& \left(
\begin{array}{c}
\nu _{\xi L}^{\prime } \\
\xi ^{\prime }
\end{array}
\right) = U\left(
\begin{array}{c}
\nu _{\xi L} \\
\xi
\end{array}\right) =\frac{1}{\sqrt{1+\frac{\left\vert \upsilon _{+}\right\vert ^{2}}{\upsilon ^{2}}}}\cdot   \label{3.13}\\
&& \left(
\begin{array}{cc}
\frac{\upsilon _{+}}{\upsilon }e^{i\theta _{\lambda _{-}}} & \frac{\upsilon
_{+}}{\upsilon }e^{i\theta _{\lambda _{+}}} \\
\frac{e^{i\theta _{\lambda _{-}}}}{\sqrt{\frac{4}{(1-\sqrt{1+4\frac{\left\vert \upsilon _{+}\right\vert ^{2}}{\upsilon ^{2}}})^{2}}}} & \frac{e^{i\theta _{\lambda _{+}}}}{\sqrt{\frac{4}{(1+\sqrt{1+4\frac{\left\vert
\upsilon _{+}\right\vert ^{2}}{\upsilon ^{2}}})^{2}}}}
\end{array}
\right) \left(
\begin{array}{c}
\nu _{\xi L} \\
\xi
\end{array}\right). \notag
\end{eqnarray}
Further making a transformation on $\nu _{\xi L}^{\prime }$, we have
\begin{equation}
\nu _{\xi L}^{\prime \prime}=i\gamma ^{5}\nu _{\xi L}^{\prime }, ~\bar{\nu} _{\xi L}^{\prime \prime} = \nu _{\xi L}^{\dagger \prime}(-i \gamma^{5})\gamma^{0} = \bar{\nu _{\xi L}^{\prime}}i \gamma^{5}.  \label{3.14}
\end{equation}
Consequently, the Lagrangian of Eq.(\ref{3.4}) can be rewritten in the standard mass-state form,
\begin{eqnarray}
\mathcal{L}_{Yukawa} & = & - \frac{m_{\xi }}{2} \left( \sqrt{ 1 + 4\frac{\left\vert \upsilon_{+}\right\vert ^{2}}{\upsilon ^{2}}} - 1 \right) \bar{\nu}_{\xi L}^{\prime
\prime}\nu_{\xi L}^{\prime \prime}  \notag\\
&& -\frac{m_{\xi }}{2}\left( 1 + \sqrt{ 1 + 4 \frac{\left\vert \upsilon_{+} \right\vert ^{2}}{\upsilon ^{2}}}\right)\bar{\xi}^{\prime}\xi ^{\prime} \label{3.15}\\
&& - \frac{m_{\xi}}{\upsilon}H\bar{\xi}\xi, ~ (\xi =e,\mu ,\tau). \notag
\end{eqnarray}

\section{Relations between masses of lepton}

Eq.(\ref{3.15}) and the leptonic dynamic terms same as in the standard model together make up the total Lagrangian relevant to leptons, not losing generality, neglecting the interaction terms with Higgs field $H$,
\begin{eqnarray}
\mathcal{L}_{Yukawa} = && \bar{\nu}_{\xi L}^{\prime \prime }i\gamma ^{\mu}\partial _{\mu }\nu_{\xi L}^{\prime \prime }+\bar{\xi }^{\prime}i\gamma ^{\mu }\partial _{\mu}\xi ^{\prime} \notag\\
&& -\frac{m_{\xi}}{2}\left(\sqrt{1+4\frac{\left\vert \upsilon_{+}\right\vert ^{2}}{\upsilon^{2}}} - 1 \right)\bar{\nu}_{\xi L}^{\prime \prime}\nu_{\xi L}^{\prime \prime} \label{4.1}\\
&& -\frac{m_{\xi }}{2}\left(1+\sqrt{1+4\frac{\left\vert \upsilon _{+}\right\vert ^{2}}{\upsilon ^{2}}}\right)\bar{\xi }^{\prime}\xi ^{\prime}, ~(\xi =e,\mu ,\tau). \notag
\end{eqnarray}
Then the Euler-Lagrange equations can be found,
\begin{equation}
i\gamma^{\mu}\partial _{\mu }\nu_{\xi L}^{\prime \prime}-\frac{m_{\xi }}{2} \left(\sqrt{1+4\frac{\left\vert \upsilon_{+} \right\vert ^{2}}{\upsilon ^{2}}}
-1 \right) \nu_{\xi L}^{\prime \prime}=0,  \label{4.2}
\end{equation}
\begin{equation}
i\gamma ^{\mu}\partial _{\mu }\xi ^{\prime} - \frac{m_{\xi }}{2} \left( 1 + \sqrt{1 + 4 \frac{\left\vert \upsilon_{+} \right\vert ^{2}}{\upsilon ^{2}}} \right) \xi^{\prime}=0, \label{4.3}
\end{equation}
($\xi =e,\mu ,\tau$). Thus, we can read the masses of leptons,
\begin{equation}
m_{\nu _{\xi L} ^{\prime \prime }} = \frac{m_{\xi }}{2}\left(\sqrt{1+4\frac{\left\vert \upsilon _{+}\right\vert ^{2}}{\upsilon ^{2}}}-1 \right), ~(\xi =e,\mu ,\tau), \label{4.4}
\end{equation}
\begin{equation}
m_{\xi ^{\prime }} = \frac{m_{\xi }}{2}\left(1+\sqrt{1+4\frac{\left\vert \upsilon
_{+}\right\vert ^{2}}{\upsilon ^{2}}}\right), ~(\xi =e,\mu ,\tau).  \label{4.5}
\end{equation}
In the modified Higgs mechanism, $\upsilon_{+}$ is assumed to be very small compared to the standard Higgs VEV $\upsilon$. Then Eq.(\ref{4.4}) and Eq.(\ref{4.5}) can be expanded in small quantity,
\begin{equation}
m_{\nu _{\xi L} ^{\prime \prime}} = m_{\xi }\frac{\left\vert \upsilon
_{+}\right\vert ^{2}}{\upsilon ^{2}}, ~(\xi =e,\mu ,\tau),  \label{4.6}
\end{equation}
\begin{equation}
m_{\xi^{\prime}}=m_{\xi }\left( 1 + \frac{\left\vert \upsilon_{+}\right\vert ^{2}}{\upsilon ^{2}}\right), ~(\xi =e,\mu ,\tau).  \label{4.7}
\end{equation}
By Eq.(\ref{4.4}) and Eq.(\ref{4.5}), we can also get
\begin{equation}
m_{\nu _{\xi L} ^{\prime \prime}} = m_{\xi ^{\prime }}\frac{\sqrt{1+4\frac{\left\vert \upsilon _{+}\right\vert ^{2}}{\upsilon ^{2}}}-1}{\sqrt{1+4\frac{\left\vert \upsilon _{+}\right\vert ^{2}}{\upsilon ^{2}}}+1}, ~ (\xi =e,\mu,\tau).  \label{4.8}
\end{equation}
This is a relation of neutrino masses depending on the known value of the  charged lepton masses, with only a single new parameter originating from our newly proposed modified Higgs mechanism.

\section{Neutrino masses Consistent with latest experimental constraints}

We show that the value of the single new parameter in the modified Higgs mechanism can be determined by the current experimental data on the neutrino mass-squared differences by using Eq.(\ref{4.8}) so that the exact value of the neutrino masses could eventually be predicted. To fulfil this purpose, performing the neutrino mass-squared differences on Eq.(\ref{4.8}),
\begin{eqnarray}
&& \Delta M_{21}^{2} = m_{\nu _{\mu L}^{\prime \prime }}^{2}-m_{\nu _{eL}^{\prime
\prime }}^{2}= \label{5.1}\\
&& \left[ m_{\mu }^{\prime }\frac{\sqrt{1+4\frac{\left\vert \upsilon_{+}\right\vert ^{2}}{\upsilon ^{2}}}-1}{\sqrt{1+4\frac{\left\vert \upsilon_{+}\right\vert ^{2}}{\upsilon^{2}}} + 1} \right]^{2} - \left[m_{e}^{\prime }\frac{\sqrt{1 + 4 \frac{\left\vert \upsilon_{+}\right\vert ^{2}}{\upsilon ^{2}}} - 1}{\sqrt{1 + 4 \frac{\left\vert \upsilon_{+}\right\vert ^{2}}{\upsilon ^{2}}} + 1 }\right]^{2}, \notag
\end{eqnarray}
\begin{eqnarray}
&&  \Delta M_{32}^{2}=m_{\nu_{\tau L}^{\prime \prime}}^{2} - m_{\nu_{\mu L}^{\prime \prime}}^{2}= \label{5.2}\\
&&  \left\lbrack m_{\tau}^{\prime}\frac{\sqrt{1+4\frac{\left\vert \upsilon_{+}\right\vert ^{2}}{\upsilon ^{2}}}-1}{\sqrt{1+4\frac{\left\vert \upsilon_{+}\right\vert ^{2}}{\upsilon ^{2}}}+1}\right]^{2} - \left[m_{\mu }^{\prime}\frac{\sqrt{1+4\frac{\left\vert \upsilon_{+} \right\vert ^{2}}{\upsilon ^{2}}} - 1 }{\sqrt{ 1 + 4 \frac{\left\vert \upsilon_{+} \right\vert ^{2}}{\upsilon ^{2}}} + 1 } \right]^{2}. \notag
\end{eqnarray}
Slightly massaging Eq.(\ref{5.1}),
\begin{eqnarray*}
\Delta M_{21}^{2} &=& (m_{\mu }^{\prime 2}-m_{e}^{\prime 2})\left[\frac{\sqrt{1+4
\frac{\left\vert \upsilon _{+}\right\vert ^{2}}{\upsilon ^{2}}}-1}{\sqrt{1+4
\frac{\left\vert \upsilon _{+}\right\vert ^{2}}{\upsilon ^{2}}}+1}\right]^{2} \\
&=& (m_{\mu }^{\prime 2} - m_{e}^{\prime 2})\left[1-\frac{2}{\sqrt{1+4\frac{\left\vert \upsilon _{+}\right\vert ^{2}}{\upsilon ^{2}}}+1}\right]^{2},
\end{eqnarray*}
gives the parameter of the vacuum perturbation,
\begin{equation}
\frac{\left\vert \upsilon_{+}\right\vert ^{2}}{\upsilon ^{2}} = \frac{1}{4} \left[ \left(\frac{2}{1-\sqrt{\frac{\Delta M_{21}^{2}}{(m_{\mu }^{\prime 2}-m_{e}^{\prime 2})}}} - 1 \right)^{2} - 1 \right].  \label{5.3}
\end{equation}
Similarly massaging Eq.(\ref{5.2}) gives,
\begin{equation}
\frac{\left\vert \upsilon_{+} \right\vert ^{2}}{\upsilon ^{2}} = \frac{1}{4} \left[ \left( \frac{2}{1-\sqrt{\frac{\Delta M_{32}^{2}}{m_{\tau }^{\prime 2} - m_{\mu
}^{\prime 2}}}} - 1 \right)^{2}-1 \right].  \label{5.4}
\end{equation}
We cite here the current experimental data on the charged lepton masses and the neutrino mass-squared differences \cite{pdg2018, Olive}, 
\begin{eqnarray}
m_{e}^{\prime } &=& 0.5109989461 \times 10^{6}~eV,  \notag \\
m_{\mu }^{\prime } &=& 105.6583745 \times 10^{6}~eV,  \label{5.6} \\
m_{\tau }^{\prime } &=& 1776.86 \times 10^{6}~eV,  \notag
\end{eqnarray}
\begin{eqnarray}
\Delta M_{21}^{2} &=& 7.53 \times 10^{-5}~eV^{2},  \label{5.7}\\
\Delta M_{32}^{2} &=& 2.54 \times 10^{-3}~eV^{2}.  \notag
\end{eqnarray}
Substituting these data into Eq.(\ref{5.3}) and Eq.(\ref{5.4}) gives,
\begin{eqnarray}
\left. \frac{\left\vert \upsilon_{+}\right\vert ^{2}}{\upsilon ^{2}} \right\vert_{\Delta M_{21}^{2}} \approx 8.21294 \times 10^{-11}, \\
\left. \frac{\left\vert \upsilon_{+}\right\vert ^{2}}{\upsilon ^{2}} \right\vert_{\Delta M_{32}^{2}} \approx 2.84140 \times 10^{-11}.
\end{eqnarray}
They agree with each other in the order of magnitude and are satisfactorily close in the numerics, at the early stage of our theoretical attempt, and also at the consideration of inevitable experimental errors. We thus adopt the average of them in the following calculations.
\begin{eqnarray}
\frac{\left\vert \upsilon _{+}\right\vert ^{2}}{\upsilon ^{2}} & = & \frac{1}{2}(8.21294 \times 10^{-11} + 2.84140 \times 10^{-11}) \notag\\
& \approx & 5.52717 \times 10^{-11}. \label{5.5}
\end{eqnarray}
Here we see that the new infinitesimal parameter we introduce to express the additional Higgs VEV is verifed by experimental data to be minuscule, with magnitude only a millionth of tha of the traditional Higgs VEV. Then, it is mathematically safe to Taylor expand Eq.(\ref{4.8}) and keep only the first order in the tiny ratio $\frac{\left\vert \upsilon _{+}\right\vert ^{2}}{\upsilon ^{2}} $ to get
\begin{eqnarray}
m_{\nu _{\xi L} ^{\prime \prime}} & = & m_{\xi ^{\prime }}\frac{\sqrt{1+4\frac{\left\vert \upsilon _{+}\right\vert ^{2}}{\upsilon ^{2}}}-1}{\sqrt{1+4\frac{\left\vert \upsilon _{+}\right\vert ^{2}}{\upsilon ^{2}}}+1} \notag\\
& \approx & m_{\xi ^{\prime}} \frac{\left\vert \upsilon _{+}\right\vert ^{2}}{\upsilon ^{2}} , ~ (\xi =e,\mu,\tau).  \label{5.6}
\end{eqnarray}
We achieve a common ratio between the neutrino mass and the mass of its corresponding charged lepton. Using Eq. (\ref{5.5}), we list our final prediction of the neutrino masses:
\begin{eqnarray}
m_{\nu _{eL} ^{\prime \prime}} & =& 2.82438 \times 10^{-5}~eV,  \label{5.8}\\
m_{\nu _{\mu L} ^{\prime \prime}} & =& 583.992 \times 10^{-5}~eV,  \label{5.9}\\
m_{\nu _{\tau L} ^{\prime \prime}} & =& 9821.01 \times 10^{-5}~eV.  \label{5.10}
\end{eqnarray}
The summation of three-generation mass values is,
\begin{equation}
\Sigma_{(\nu)} ~ m_{\nu} = m_{\nu _{eL}^{\prime \prime}} + m_{\nu _{\mu L}^{\prime \prime}} + m_{\nu _{\tau L}^{\prime \prime}} \approx 0.104078 ~eV.\label{5.11}
\end{equation}
This result satisfies the constraint from the current comsmological observations \cite{Giusarma}, which we cite here:

``If high-multipole polarization data from Planck is also considered, the 95\% C.L. upper bound is tightened to
\begin{equation}
``\Sigma_{(\nu)} ~ m_{\nu} < 0.176 ~eV."  \label{5.12}
\end{equation}
This consistency presents a very positive signal to the soundness of our proposal of neutrino mass generation. We claim to have obtained the long-dreamed exact mass values for all three generations of neutrinos that satisfy the constraints of current physical experiments.

\section{Conclusion and Discussion}

Symmetry has been found a fundamental principle in the theory of particle physics. But afterwards many of the symmetries in theory have been found broken in reality, or had to be so to account for the physics in reality. The so-called Higgs mechanism proposes a new scalar field with symmetry breaking in the vacuum state to interact with fermions and non-abelian gauge bosons so that the massiveness of non-abelian gauge bosons and most of the fermions receives excellent explanation. The standard electroweak model is then rendered a practically complete structure. However, the massiveness of neutrinos only was generally realized after the completion of the standard model and is not explained by the standard model.

This work proposes a modified Higgs mechanism by extending the standard Higgs field vacuum breaking to one with an extra but tiny breaking in the vacuum of the charged Higgs field. This modification to the standard Higgs mechanism provides an simplistic, natural and physical inspiring explanation for the mechanism of neutrino mass generation. We show that this mofied Higgs mechanism when applied to the sector of neutrinos can produce very desirable results for neutrino masses. The modified Higgs mechanism should also be applied to gaugue bosons and all other particle sectors that interact with Higgs field. It can be qualitatively conjectured and also it can be shown that the modified Higgs mechanism with apparently QED breaking causes only negligible corrections to the masses of particles which are already massive in standard Higgs mechanism, and fortunately also negligible correction to the photon mass, a correction that is well below the constraint determined by latest experiments so that an apparently QED breaking proposal does not really break QED in real physics. The technical proof of this problem and more other applications of the modified Higgs mechanism will be arranged into independent articles that our group will release soon.

The only new independent parameter we introduce is a quantity parametrizing the additional Higgs vacuum breaking, in a way compatible with current physical experiments. The validity of the relevant new physics is to be verified in the physical experiments or cosmological observations. The fact that the neutrino masses we predict in this work by using the particle experimental data could sum up to satisfy the constraint of current cosmological observations shows very positive signal to the soundness of this new idea of neutrino physics.

The effectiveness, simplicity and naturalness of our scheme of neutrino mass generation could be considered as solidly reasonable due to three reasons: firstly, the tiny vacuum breaking parameter is the only new parameter of our model; secondly, our model assigns a common Yukawa coupling constant to the neutrino and its corresponding charged lepton in the same family rather than introducing one separate Yukawa coupling constants for each of them; and thirdly, the physical picture of our model originates from natural extension of the standard Higgs mechanism.

In summary, in this article we have proposed a two-step vacuum breaking Higgs mechanism in which the Higgs vacua break twice. Maintaining the identical particle constituents and the full set theoretical symmetries of the standard electroweak model, the doubly-vacuum-breaking Higgs mechanism provides us with very simplistic explanation of the neutrino mass generation. In the view of the doubly-vacuum-breaking Higgs mechanism, the very minuteness of the charged Higgs vacuum breaking explains the origin of the very minuteness of neutrino masses, while the relative greatness of the standard Higgs vacuum breaking is recognized as the origin of the relative greatness of charged lepton masses. This proposal can bring rich new physics to Higgs-relevant problems in particle physics, besides injecting new inspiration to the solution of massive neutrino problems.

\begin{acknowledgments}
We would like to thank Dr. Z. H. Xiong and Dr. W. Y. Wang for their useful and interesting discussion. This work is partly supported by NSF through a CAREER Award PHY-0952630, by DOE through Grant DE-SC0007884 in USA, and by the National Natural Science Foundation of China (Grants No. 11875081 and No. 11173028).
\end{acknowledgments}

\end{document}